\documentclass[showpacs,twocolumn,preprintnumbers,amsmath,amssymb,prl,epsf,superscriptaddress]{revtex4-1}

\usepackage{graphicx}
\usepackage{dcolumn}
\usepackage{bm}
\usepackage{hyperref}
\usepackage{amsmath,amssymb}
\usepackage{physics}

\usepackage{mathrsfs}
\usepackage{mhchem}
\usepackage{siunitx}
\usepackage{soul, xcolor}

\setstcolor{violet}

\begin{document}


\title{Microscale generation of entangled photons without momentum conservation}

\author{C. Okoth}
\email{cameron.okoth@mpl.mpg.de}
\affiliation{Max Planck Institute for the Science of Light, Staudtstra{\ss}e 2, 91058 Erlangen, Germany}
\affiliation{University of Erlangen-N\"urnberg, Staudtstra{\ss}e 7/B2, 91058 Erlangen, Germany}
\author{A. Cavanna}
\affiliation{Max Planck Institute for the Science of Light, Staudtstra{\ss}e 2, 91058 Erlangen, Germany}
\affiliation{University of Erlangen-N\"urnberg, Staudtstra{\ss}e 7/B2, 91058 Erlangen, Germany}
\author{T. Santiago-Cruz}
\affiliation{University of Erlangen-N\"urnberg, Staudtstra{\ss}e 7/B2, 91058 Erlangen, Germany}
\affiliation{Max Planck Institute for the Science of Light, Staudtstra{\ss}e 2, 91058 Erlangen, Germany}
\author{M. Chekhova}
\affiliation{Max Planck Institute for the Science of Light, Staudtstra{\ss}e 2, 91058 Erlangen, Germany}
\affiliation{University of Erlangen-N\"urnberg, Staudtstra{\ss}e 7/B2, 91058 Erlangen, Germany}
\affiliation{Department of Physics, M. V. Lomonosov Moscow State University, Leninskie Gory, 119991 Moscow, Russia} 

\date{\today}
\maketitle

\textbf{Spontaneous parametric down-conversion (SPDC) is a well-developed tool to produce entangled photons for practical applications, such as quantum imaging \cite{lemos2014quantum,pittman1995optical,brida2010experimental}, quantum key distribution \cite{adachi2007simple,jennewein2000quantum} and quantum metrology \cite{wu1986generation,valencia2004distant}, as well as for tests of quantum mechanics \cite{shih1988new}. So far, SPDC sources have exclusively been used in the phase matched regime: when the emitted daughter photons conserve the momentum of the pump photon. Phase matched SPDC restricts both the choice of non-linear materials and the available states the daughter photons can occupy. This has prompted the search for SPDC where phase matching is absent \cite{saleh2018towards,tokman2018purcell,yamashita2018nonlinear,marino2019spontaneous}. Here we register over 1000 photon pairs per second generated in a micrometer-sized SPDC source continuously pumped at a moderate power. This is both the smallest SPDC source reported to date and, due to the relaxed phase matching, an entirely new type of two-photon radiation. Reducing the interaction length and localizing the position of the daughter photons creation leads to a huge uncertainty in their momentum. The uncertain momentum mismatch leads to a frequency-angular spectrum an order of magnitude broader than that of phase matched SPDC. This results in record breaking spatio-temporal correlation widths and huge entanglement. We believe that similar sources can be realized on thinner platforms, not only compounding the aforementioned effects, but also allowing easy integrability on optical chips.}

For SPDC, the probability of a pump photon, with momentum $\hbar \vec{k}_p$ (wavevector $\vec{k}_p$), to decay into two daughter
photons, signal and idler with momenta $\hbar \vec{k}_{s,i}$ (wavevectors $\vec{k}_{s,i}$), strongly depends on the momentum (wavevector) mismatch $\hbar \vec{\Delta k}\equiv \hbar \vec{k}_s+\hbar \vec{k}_i-\hbar \vec{k}_p$. The allowed mismatch forms an uncertainty relation with the volume over which the non-linear interaction takes place. Its component parallel to the pump, or \textit{longitudinal mismatch} $\Delta k_{||}$, is restricted by the inverse length of the nonlinear material. The component perpendicular to the pump, or \textit{transverse mismatch} $\Delta k_{\bot}$, is restricted by the inverse non-linear interaction area, which is generally given by the Gaussian profile of the pump beam.
  
In a phase-matched process, $\Delta k_{||}=0$, the longitudinal momentum of the pump photon is conserved by the daughter photons. However, this is a special case and is only satisfied in a few non-linear materials with the appropriate electronic and optical properties. More generally $\Delta k_{||} \neq 0$, causing pairs to be generated in or out of phase with respect to pairs generated earlier in the non-linear material. The relative phase difference depends on the non-linear interaction length, $L$.  If $L$ is equal to an odd multiple of the so called coherence length, $L_{c} = \frac{\pi}{\Delta k_{||}}$~\cite{boyd2003nonlinear},  the interference between the emitted pairs is fully constructive leading to a local maximum in the emission probability. At $L<L_c$, the pairs are always generated in phase. Moreover, with small length $L$, the allowed longitudinal mismatch can be very large, which leads to a very broad spectrum of emitted photons, both in frequency and in angle. 
\begin{figure}
\includegraphics[width=\columnwidth]{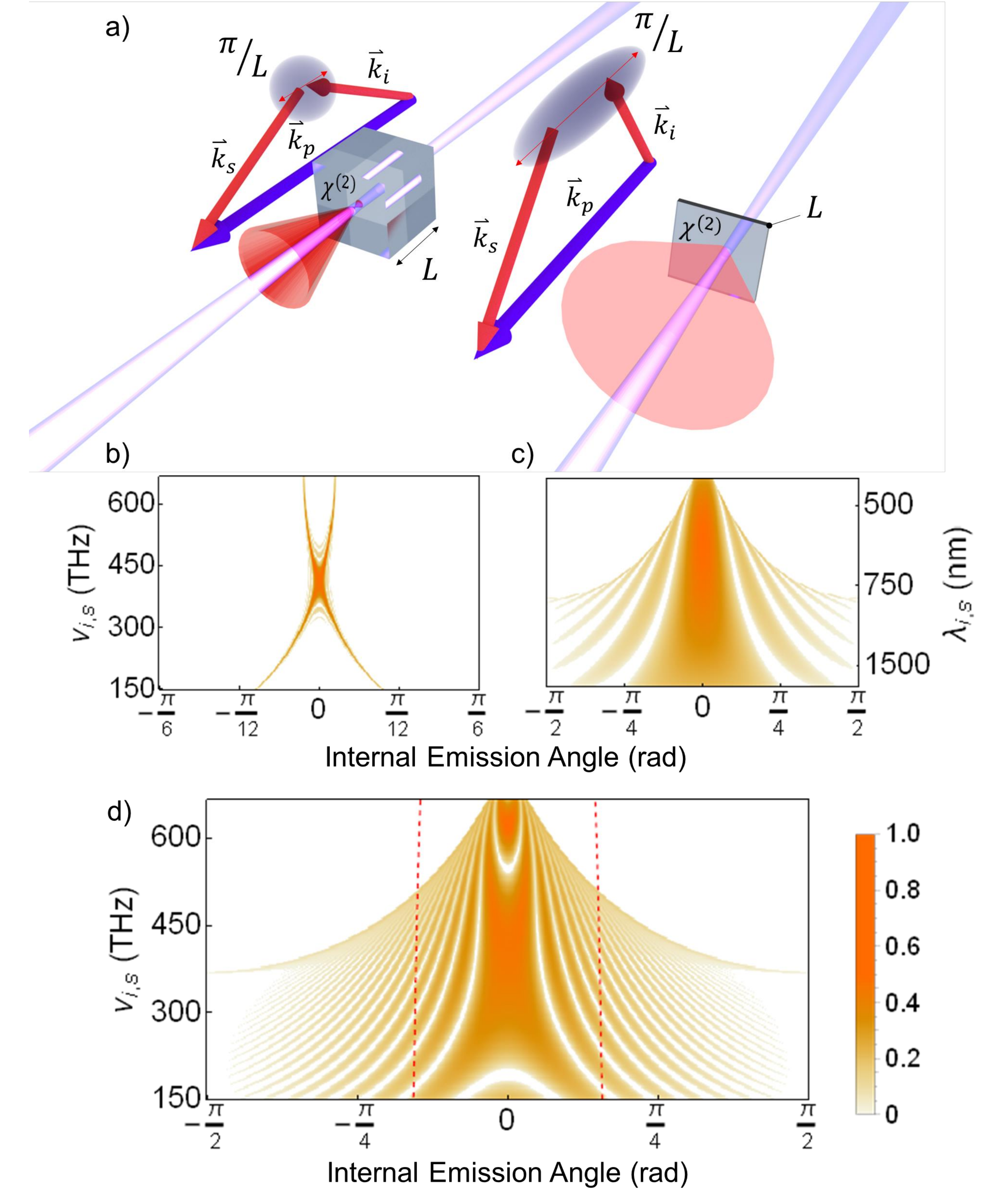} 
\caption{(a) SPDC in a thick (left) and ultrathin (right) nonlinear layer orthogonal to the pump. In the second case, the allowed momentum mismatch along the pump direction is very large, which results in a broad spectrum, both in  frequency and angle. As an example, the middle panels show the frequency-angular spectrum calculated for a phase matched type-I BBO crystal with $L=1$ mm (b) and non-phase matched type-0 lithium niobate crystal with $L= L_c= 1.37 \mu$m (c). The pump wavelength is $405$ nm and beam waist $100\,\mu$m. Note the different x-axis scales between (b) and (c). Panel (d) is the frequency-angular spectrum expected for a non-phase matched type-0 lithium niobate crystal with $L= 5 L_c$.  The red dashed lines show the angle of internal reflection, which limits the angle collected.}
\label{fig:phasematchingandnonphasematching}
\end{figure} 

Fig.~\ref{fig:phasematchingandnonphasematching} a shows the allowed mismatch for a thick crystal and an ultra-thin layer. Whilst the thick crystal restricts the signal and idler angles and frequencies, the ultra-thin layer allows a broad range of modes to be populated. The calculated frequency - angular spectrum for SPDC from a single coherence length or its odd multiple (Fig.~\ref{fig:phasematchingandnonphasematching} c,d) is far broader than any phase matched SPDC spectra observed from a macroscopically thick crystal (Fig.~\ref{fig:phasematchingandnonphasematching} b). The emission characteristics differ so greatly between phase matched and non-phase matched SPDC that they can be considered separate sources of photon pairs \cite{yuan2019spatiotemporally}.

While the possible emission angles and frequencies for the signal and idler photons are determined by the longitudinal mismatch, the \textit{correlation} between the signal and idler angles of emission is governed by the transverse mismatch~\cite{Burlakov1997,Monken1998}. These correlations will be tight if the pump has a large beam waist. Similarly, the frequency correlations are governed by the bandwidth of the pump~\cite{Keller1997,Grice1997} and will be tight for a narrowband continuous-wave pump. For this reason, SPDC from an ultra-thin layer should produce photon pairs that are highly entangled in angle and frequency: while these parameters are very uncertain for a single photon, they are known with certainty when the conjugate photon is detected. It is noteworthy that whilst previous attempts to boost the bipartite entanglement have primarily focused on increasing either frequency or angular entanglement~\cite{Fedorov2007,Brida2009}, SPDC in an ultrathin layer {\it simultaneously} has gains in both, leading to a massive improvement over previous techniques.   

The reduced interaction length, however, leads to a lower SPDC emission probability (\ref{eq:probability}). This can be partly compensated for by using a highly nonlinear medium. Here we use an x-cut lithium niobate (LN) crystal and, by taking advantage of the fact that we no longer need to phase match, utilize the large $d_{33}$ component of the nonlinear susceptibility tensor by correctly orienting the pump polarization. The $d_{33}$ component of LN ($40$ pm/V) is roughly 40 times stronger than the effective susceptibility of $\beta-$Barium borate (BBO), a standard crystal used for SPDC experiments \cite{boyd2003nonlinear}.

\begin{figure}
\includegraphics[width=\columnwidth,clip]{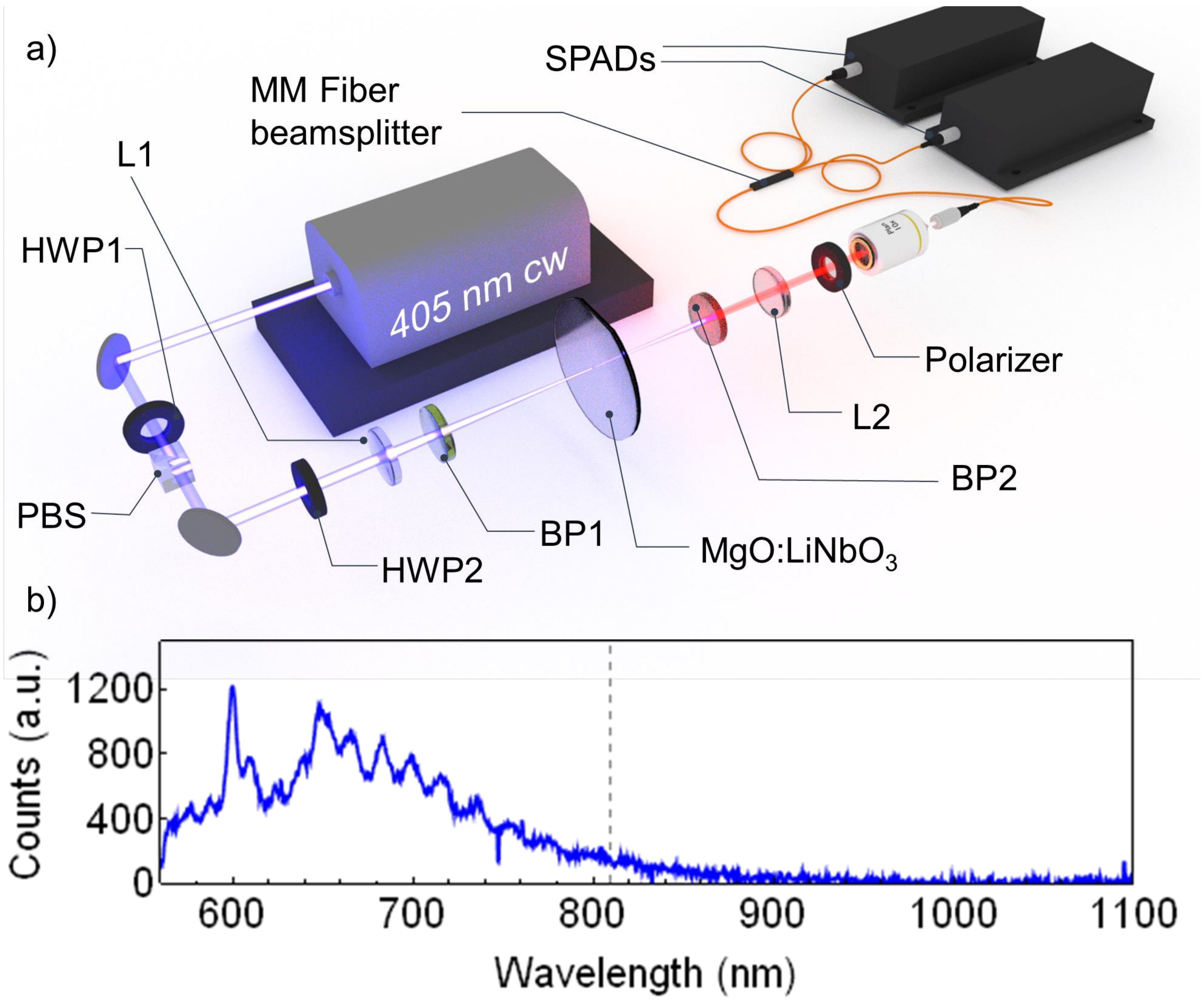}
\caption{ The setup used to detect SPDC from LN (a) and the fluorescence spectrum measured  with a spectrometer (b). The dashed line indicates the degenerate wavelength where most of the measurements were taken.}
\label{fig:setup}
\end{figure}

The experimental setup is shown in Fig.~\ref{fig:setup}. The sample tested was a thin layer of magnesium oxide doped LN on a $500\,\mu$m fused silica substrate. The $z$-axis of LN was in the plane of the layer. The thickness of the sample varied from 5.9 $\mu$m to 6.8 $\mu$m, due to the non-uniform fabrication of the wafer. Therefore by scanning the LN in the z-y plane it was possible to tune the length.

Despite the use of the highest non-linear component available in LN, the efficiency remained much lower than for phase-matched SPDC in a macroscopic crystal. As a result, fluorescence, which can usually be disregarded when working with phase matched SPDC, became the dominant process in the spectral region of interest. The measured fluorescence (Fig.~\ref{fig:setup} b) was more than an order of magnitude stronger than the expected SPDC emission. Due to the broadband nature of non-phase matched SPDC, distinguishing the two-photon radiation from fluorescence could only be done using correlation measurements.

\begin{figure}
\includegraphics[width=8.9cm]{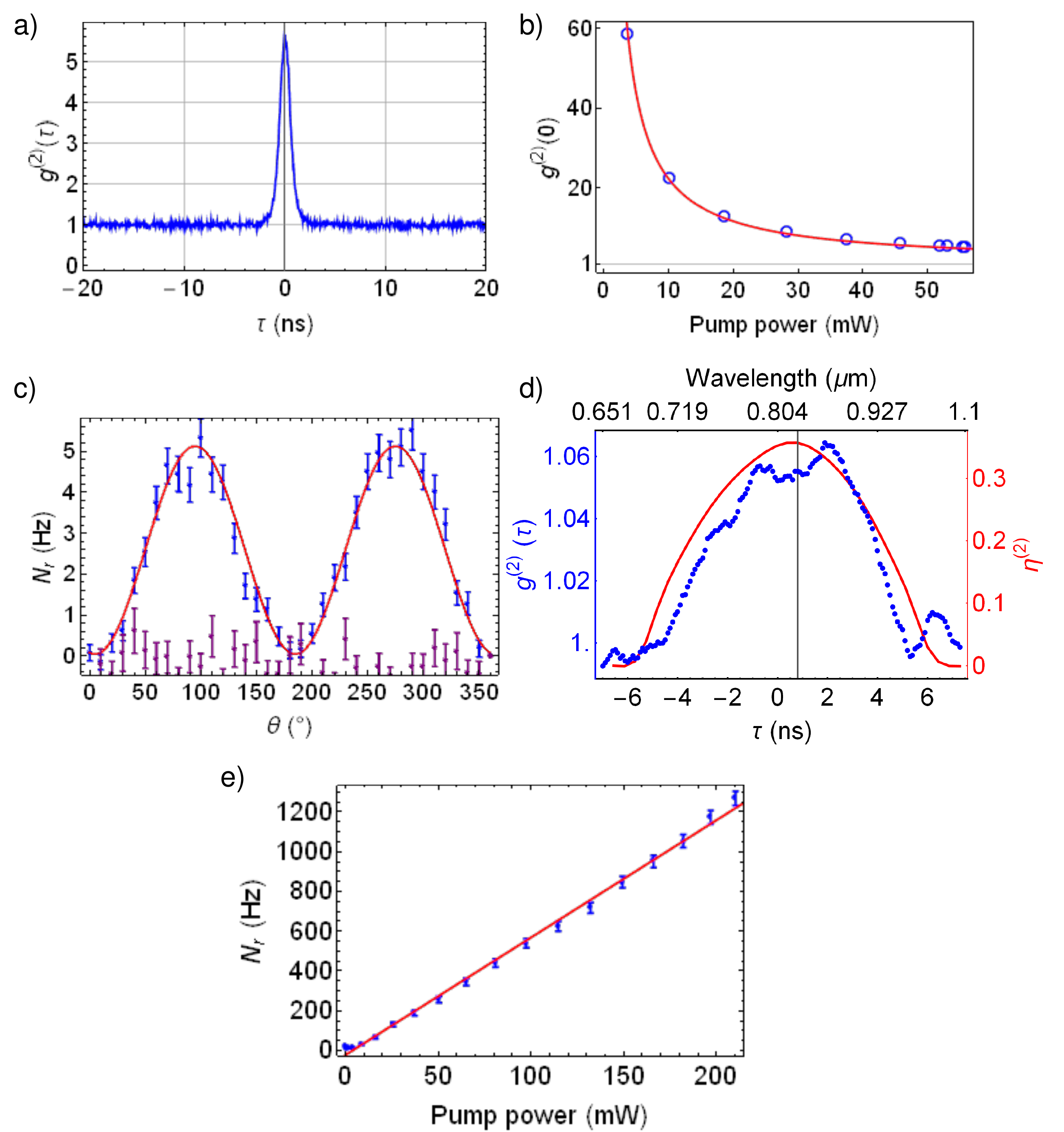} 
\caption{Two-photon correlations from the LN sample. (a) The  normalized two-photon temporal correlation function. (b) The normalised correlation function at zero time delay measured  versus the pump power (blue points) and fitted with an inverse dependence (red line). (c) The rate of real coincidences versus the pump polarisation direction, for the emission polarised along z (blue points, red fit) and along y (purple points). (d) The normalized temporal correlation function measured after propagation through a dispersive fiber. The temporal delay is mapped into units of wavelength. The double photon quantum efficiency, which limits the collection bandwidth (see Methods), is shown in red. The gray line indicates the degenerate frequency. (e) The coincidence rate (blue points), for optimised photon pair collection, as a function of the pump power centered at 500 nm with the expected linear power dependence (red).}
\label{fig:SPS}
\end{figure}

The correlation measurements were performed using Hanbury Brown - Twiss (HBT) setup (Fig.~\ref{fig:setup} a). To begin with, the normalized second-order correlation function $g^{(2)}(\tau)$ was {measured (see Methods),} shown in Fig.~\ref{fig:SPS} a. Strong two-photon correlations were observed, as we see from the peak at zero delay. Further, the value of $g^{(2)}(0)$ was measured for different varied pump powers (see Fig.~\ref{fig:SPS} b). The measured value of $g^{(2)}(0)-1$ was shown to have an inverse dependence on the pump power, which is a fingerprint of two-photon emission ~\cite{ivanova2006multiphoton}.  The correlation function $g^{(2)}(0)$, related to the coincidences-to-accidentals ratio $\texttt{CAR}=g^{(2)}(0)-1$, had a high value and an inverse dependence on the mean number of photons, both of which are compelling evidence of two-photon light generation.

The measured polarization dependence of the coincidence count rate (Fig.~\ref{fig:SPS} c) confirms that the SPDC was mediated by the $d_{33}$ component of the nonlinear susceptibility tensor, with the pump, signal, and idler photons all polarized along the z axis. Indeed, the SPDC was noticeable only when the signal and idler photons were polarized along the z axis, in this case the coincidence rate depended on the angle $\theta$ between the pump polarization and the y axis as $\sin^{2}{(\theta)}$ (blue points, red fit). For the emission polarized along the y-axis, no real coincidences were observed (purple points).



The SPDC spectrum was measured using single-photon spectroscopy (SPS) \cite{valencia2002entangled,avenhaus2009fiber} (see Methods), which allowed us to distinguish it from the fluorescence spectrum. Light generated from LN was coupled into a dispersive fiber. After propagation through the fiber the two-photon wavepackets spread in time. Therefore, the difference in the signal and idler photon detection times could be mapped to a frequency difference. The measured normalized correlation function after the propagation through 160 m of optical fibre is shown in Fig.~\ref{fig:SPS} d. The width of the peak corresponds to 200 nm. This was less than the expected width of $600$ nm (Fig.~\ref{fig:phasematchingandnonphasematching} d) because of the relatively narrow sensitivity range of the single-photon detectors (see Methods) and frequency-dependent fiber coupling. The biphoton correlation time, given by the inverse spectral width~\cite{valencia2002entangled}, was estimated to be $10$ fs. The full spectral range of $600$ nm corresponds to a correlation time of $3$ fs.

Measurements with a 405 nm pump yielded a modest coincidence count rate due to the high fluorescence background saturating detectors even at low powers. To optimise this we moved to a 500 nm CW pump operating in a similar power range. The shift in wavelength reduced the fluorescence allowing a broader angle of emission to be collected without saturating the detectors. The linear dependence between the coincidence rate and pump power was verified for the optimised wavelength and collection angle (see Fig.~\ref{fig:SPS} e). For the optimised pair collection and pump wavelength the CAR was measured to be 1400, with 100 mW pump power.

\begin{figure*}
\includegraphics[width=17cm]{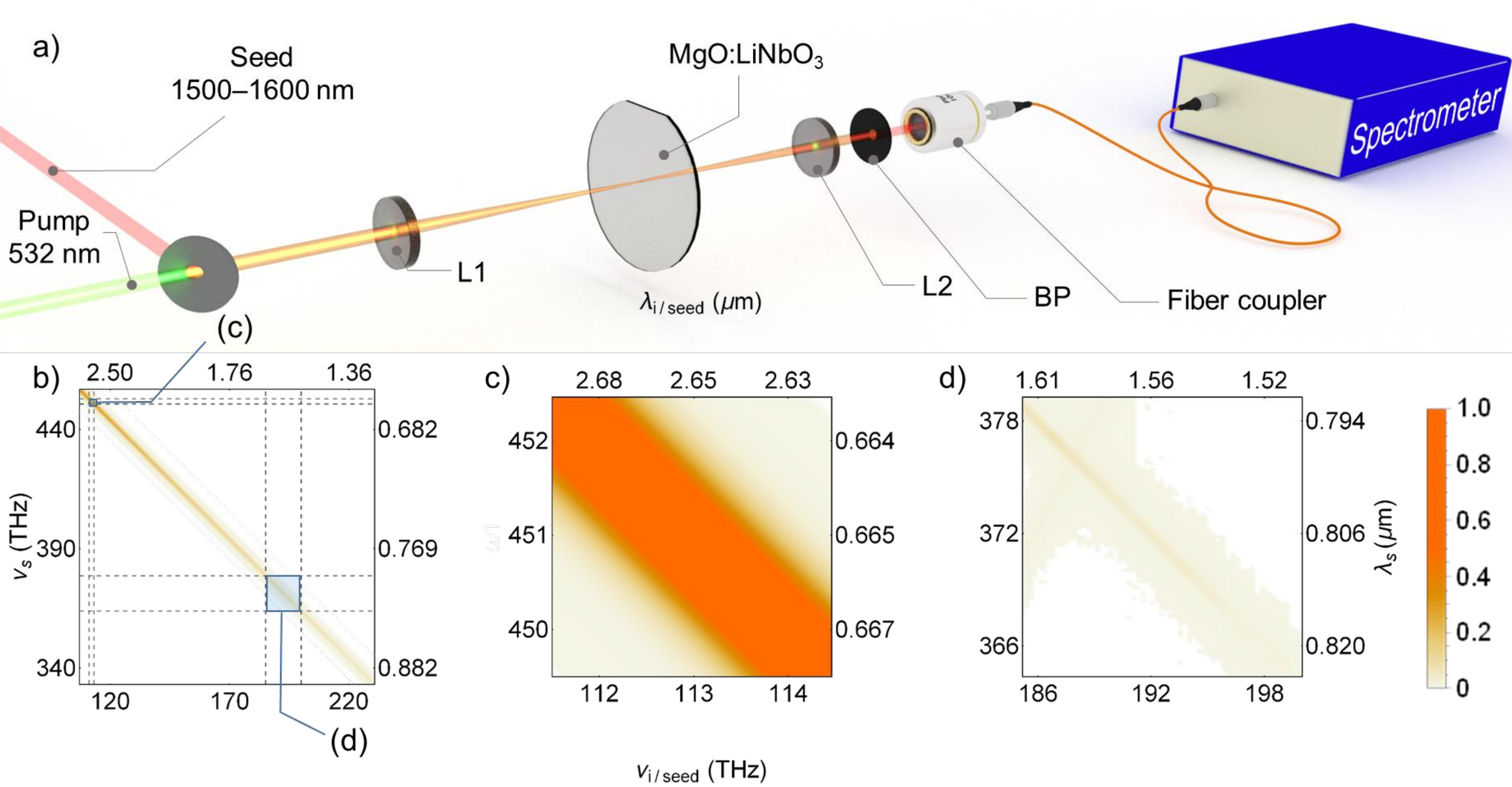} 
\caption{The setup for the SET measurement (a) and the calculated JSI for 5.8 $\mu$m LN pumped at $532$ nm (b), with a zoomed in interval (c) to emphasise the small width. L1 and L2 are lenses and BP is a band-pass filter blocking the pump and the seed. Panel (d) shows a fragment of JSI measured using SET, with the bounds limited by the tunability of the seed. The fainter line with a positive tilt in (d) is due to second harmonic generation of the seed.}
\label{fig:JSI}
\end{figure*}

To demonstrate the high degree of frequency entanglement, we measured the joint spectral intensity (JSI), quantifying the joint probability $P(\omega_s,\omega_i)$ of the signal and idler photons having frequencies $\omega_s,\omega_i$, respectively (see Methods). The JSI was reconstructed using stimulated emission tomography (SET)~\cite{liscidini2013stimulated}, see Fig.~\ref{fig:JSI} a.


By pumping with the second harmonic of the Nd:YAG laser at $532$ nm and tuning the seed beam wavelength between $1500$ and $1620$ nm, a small part of the JSI (Fig.~\ref{fig:JSI}b,c) was mapped out as shown in  Fig.~\ref{fig:JSI} d. Due to the narrow tuning range of the seed, the full spectral range of SPDC could not be obtained. Still, the SET results demonstrate tight frequency correlations ($0.6$ THz), within a $13$ THz range of the JSI, which confirms a degree of entanglement of at least $20$, a value that becomes far higher when taking into account the broad spectral width. 

Generating entangled photons with a broad spectral and angular width has been a long standing goal in the field of quantum optics \cite{nasr2008ultrabroadband,bernhard2013shaping,katamadze2015broadband}. Highly non-linear ultrathin layers provide both a platform on which to achieve this goal and, due to their size and scale, a platform on which to design miniaturized quantum-photonic chips \cite{guo2017parametric}. To our knowledge, here we have reported the smallest source of SPDC ($6 \mu m \times 10 \mu m \times 10 \mu m $) found to date. Several semiconductor materials, such as gallium arsenide, have much higher second-order susceptibilities than that of LN \cite{dmitriev2013handbook}. In addition, structuring such materials to enhance their resonance response can further improve the SPDC efficiency~\cite{Liu2018}, thus allowing even thinner samples to be made.

The two-photon time and space correlation widths, proportional to the inverse spectral and angular widths, respectively,
promise to be tighter than anything currently observed. As such the correlations in space and time could become a new resolution standard. Whilst the ultrashort correlation time can be used to synchronize distant clocks \cite{valencia2004distant}, the tight spatial correlations can dramatically improve the resolution of many quantum imaging techniques such as ghost imaging and imaging with undetected photons \cite{pittman1995optical,lemos2014quantum,brida2010experimental,d2005resolution}. In two-photon microscopy, such a source will allow imaging well beyond the diffraction limit~\cite{denk1990two}, as the resolution will be determined only by the non-phase matched SPDC interaction length. 

Both the angle of emission and frequency can be used as variables for encoding quantum information \cite{olislager2010frequency,rarity1990experimental}. Due to the huge spectral and angular width of non-phase matched SPDC, the expected degree of entanglement, and thus the information capacity, for non-phase matched SPDC is almost an order of magnitude higher than for a phase matched process. 
Additionally, using the principle of hyperentanglement, whereby the biphoton state is entangled in more than one degree of freedom \cite{barreiro2005generation}, the non-phase matched state increases the information capacity as the product of both the entanglement in the frequency and angular domain, leading to improvements by several orders of magnitude.


Finally, we expect that highly non-linear thin layers will also be suitable platforms on which to observe higher-order parametric down conversion effects, such as the generation of three-photon states from a cubic interaction \cite{okoth2019seeded}. Phase matching for higher-order processes becomes increasingly challenging due to the larger refractive index difference between the pump photon and daughter photons. This restricts the number of non-linear materials to work with to a far greater extent than for second-order processes. Therefore working in the non-phase matched regime is a reasonable trade-off as it allows one to utilise materials with a huge cubic susceptibility.


\section{\label{sec:methods} Methods}

\textit{Longitudinal and transverse mismatch.}
In a material of length $L$, with the nonlinear susceptibility $d$, the probability of a pair emission is  
\begin{equation} \label{eq:probability}
    P\propto d^{2} ~ L^2 F_{pm} (\Delta k_{||}) F_{p} (\Delta k_{\bot}),
\end{equation}
where $F_{pm} (\Delta k_{||})$ is the phase matching function, 
\begin{equation} \label{eq:sinc}
        F_{pm} (\Delta k_{||})=\text{sinc}^2 \left(\dfrac {\Delta k_{||} L}{2}\right), 
\end{equation}
and, assuming a Gaussian pump beam profile with a waist $w$, $F_p (\Delta k_{\bot})$ is the pump function, 
\begin{equation} \label{eq:Gauss}	   
F_{p} (\Delta k_{\bot})=e^{-(\Delta k_{\bot}w)^2}.
\end{equation}
The phase matching function, which depends solely on the longitudinal mismatch $\Delta k_{||}$, dictates the overall spectral and angular width of the two-photon emission if the length $L$ is not too large. Conversely, the pump function, which depends on the transverse mismatch component, $\Delta k_{\bot}$, gives the angular correlation width between the signal and idler photons. 

\textit{Scanning the sample thickness.} 
The rate of non-phase matched SPDC emission has an oscillatory dependence on the layer thickness, being maximum for odd multiples of $L_c$ and zero for even multiples of $L_c$. In the case of a $405$ nm pump and degenerate SPDC at $810$ nm the coherence length for LN is $L_c=1.37\,\mu $m. By varying the position of the pump beam on the sample, it was possible to scan its thickness (see Supplementary Information, Fig.~\ref{fig:thickness}). The maximal emission happened in the thicker region of the sample, which corresponded to about $5L_c$, as shown in Fig.~\ref{fig:2DCoincMap} of the Supplementary material. Note that the rate of SPDC for thickness $L_c$ would be the same as for thickness $5L_c$, albeit with a broader spectrum (see Fig.~\ref{fig:phasematchingandnonphasematching} c and d), but only a thicker sample was available for technical reasons.

\textit{Registering coincidences.}\label{regcoi}
Two-photon coincidences were detected using a Hanbury Brown-Twiss setup (Fig.~\ref{fig:setup} a). The CW pump beam was centered at $405$ nm and had a maximum power output of $60$ mW (Omicron). The pump polarization and power were controlled using a polarizing beam splitter (PBS) and two half wave plates (HWP). The pump was focused, using a lens (L1) with a $100$ mm focal length, into the LN sample, which was placed on a translation stage to provide movement in the z-y plane of the crystal structure. To avoid additional sources of fluorescence the pump was filtered using a 405 nm bandpass (BP1) immediately after the L1. The resulting emission was filtered of the pump with an 810 nm bandpass filter (BP2), collected using a lens (L2) with a $100$ mm focal length and coupled into a multimode fiber beam splitter. To isolate the breakdown flash effect of the single-photon detectors in time~\cite{newman1955visible}, two $20$ m long MM fibers were placed after the fiber beam splitter. The photons were registered on two Perkin-Elmer visible-range single photon avalanche diodes (SPAD). The arrival time of each photon was recorded using a Swabian Instruments Time Tagger.
To find the real coincidence rate, the accidental coincidence rate was subtracted, 
\begin{equation}
\label{eq:real}
    N_r=N_c-N_a,
\end{equation}
where $N_c$ is the total coincidence rate and $N_a=N_1 N_2 T_c$. $N_{1,2}$ are the single photon count rates into detector 1 and 2 respectively, and $T_c$ is the coincidence resolution time. The normalized second-order correlation was calculated using
\begin{equation}\label{eq:g2}
    g^{(2)}(0)=\dfrac{N_c}{N_a}.
\end{equation}
From this it is clear that $g^{(2)}(0)-1$ is inversely proportional to the number of photons detected: whilst the number of pairs scales linearly with the pump power the number of accidentals scales quadratically with the pump power.

For the optimised collection of correlated pairs the pump was switched from 405 nm to 500 nm \cite{ECD}. To accommodate the change in pump the SPAD's were swapped for single-mode superconducting nanowires with a large quantum efficiency in the near-infrared. The reduction in fluorescence meant a larger angle of emission could be collected. For more efficient coupling to the detectors, the pump was focused into a 10 micron spot size.

\textit{Single-photon spectroscopy.} \label{sps}
To measure the SPDC spectrum, we introduced a $160$ m SMF28 fibre into the setup shown in Fig.~\ref{fig:setup} a, between the LN output and the MM fiber beam splitter. L1 was replaced with a lens  that had a 20 mm focal distance to decrease the pump beam waist in an effort to lower the number of spatial modes and increase the coupling efficiency into the SMF28 fiber. Finally, a 645 nm long-pass filter was used for BP2. The dispersive temporal walk-off in the fibre led to a difference in photon arrival times corresponding to a specific frequency difference. To calibrate the arrival time differences, the radiation in one arm of the fiber beam splitter was sent through a $10$ nm broad interference filter centered at the calibration wavelength. The emission was recoupled into the $20$ m MM fibre after passing through the filter and the coincidences between the first and the second detector were measured as a function of the difference in photon arrival times. This was done for six interference filters and a calibration curve between time differences and frequency differences was fitted with a cubic polynomial (see Fig.~\ref{fig:calibration} of the Supplementary material). As this type of measurement was based on collecting correlated photons, any frequency dependent loss on one photon was mirrored on its conjugate. This affected the spectral sensitivity of the detectors {when operating as coincidence counters} which, accounting for the mirrored quantum efficiency of the near-infrared to the visible region, is shown as the red curve in Fig.~\ref{fig:SPS} d.


\textit{Stimulated emission tomography.}\label{set}
A custom optical parametric amplifier (OPA) was built using a pair of bismuth triborate (BiBO) crystals (see Fig.~\ref{fig:setupset} of the Supplementary material) with type-I phase matching. It was pumped by a frequency doubled Nd:YAG laser (532 nm) with a $20$ ps pulse width and 1 kHz repetition rate. The pump was used to amplify a tunable (1500-1620 nm) CW source. The output idler radiation of the OPA had the same pulse properties as the pump beam and was used to seed the SPDC in LN for the SET measurement. To compensate for any temporal walk-off between the pump and the seed, a delay line between the two beams was built. A PBS and HWP were added to both arms to control both the power and the polarization of the seed and the pump locally. The signal beam of the OPA was suppressed by the PBS in the pump arm and a frequency filter in the seed arm. The seed and pump beams were overlapped spatially and temporally (see Fig.~\ref{fig:JSI} a) and focused into the LN sample using a $400$ mm lens  (L1). The seed wavelength was scanned and the stimulated PDC radiation was registered with an Avantes visible-range spectrometer. The resulting spectra were plotted for each seed frequency probed to reconstruct the JSI. 

The expected JSI, in frequency space, for collinear SPDC is   
\begin{equation}
F(\omega_s,\omega_i)= F_{pm}(\omega_s,\omega_i)F_{p}^{(\omega)}(\omega_s,\omega_i),
\label{eq:JSI} 
\end{equation}
where $F_{pm}(\omega_s,\omega_i)$ is given by (\ref{eq:sinc}) with $\Delta k_{||}=\Delta k_{||}(\omega_s,\omega_i)$ and, for a Gaussian pump,
\begin{equation} \label{eq:freqpm}
        F_{p}^{(\omega)}(\omega_s,\omega_i)=e^{  -\dfrac{1}{2}\left(\dfrac{\omega_{s}+\omega_{i}-\omega_{p_0}}{ \sigma}\right)^2 },
\end{equation}
$\omega_{p_0}$ is the central pump frequency and $\sigma$ is the pump spectral width \cite{hendrych2007tunable}.

\textit{The degree of entanglement.}
Due to the lack of phase information in our experiment we estimate the degree of entanglement using a measure introduced by Fedorov~\cite{fedorov2004packet}. It is defined as the ratio $R=\Delta/\delta$, where  $\Delta$ is the unconditional width of the JSI, i.e., the spectral width of the signal (idler) radiation, and $\delta$ is the corresponding conditional (correlation) width i.e. the spectral width of the idler (signal) radiation once its conjugate has been measured. The Fedorov ratio $R$ is a valid measure under certain conditions, in particular where the so-called double Gaussian approximation is valid. We estimate, by comparing simulated phase matched and non-phase matched JSI's, that it deviates from the Schmidt number $K$~\cite{Mikhailova2008,Brida2009} by less than a factor of two for the non-phase matched regime. In Fig.~\ref{fig:JSI} c, $\delta$ is the width of the JSI cross-section, $0.6$ THz. As to $\Delta$, it was measured to be $13$ THz but it is limited by the range probed in experiment. Therefore, we can confirm $R\ge20$ but in reality, the degree of entanglement should be at least an order of magnitude larger, according to Fig.~\ref{fig:JSI} b. 

\subsection*{Acknowledgements}

We acknowledge the financial support by Deutsche Forschungsgemeinschaft (DFG) (CH-1591/3-1). 

\subsection{Additional Information}

\textbf{Author Contributions} All authors conceived, contributed and discussed the final results published in this manuscript. A. C., T. S-C. and C.O. performed the experimental work and carried out the data analysis. M.C. initiated and supervised the project.

\textbf{Data Availability} The data that support the findings of this study are available from the corresponding
author upon reasonable request.

\textbf{Competing Interests} The authors declare that there are no competing interests.

\bibliographystyle{ieeetr}
\bibliography{bibliography}

\newpage
\section{\label{sec:Supplementary} Supplementary information}

\begin{figure}[htb]
\includegraphics[width=7cm]{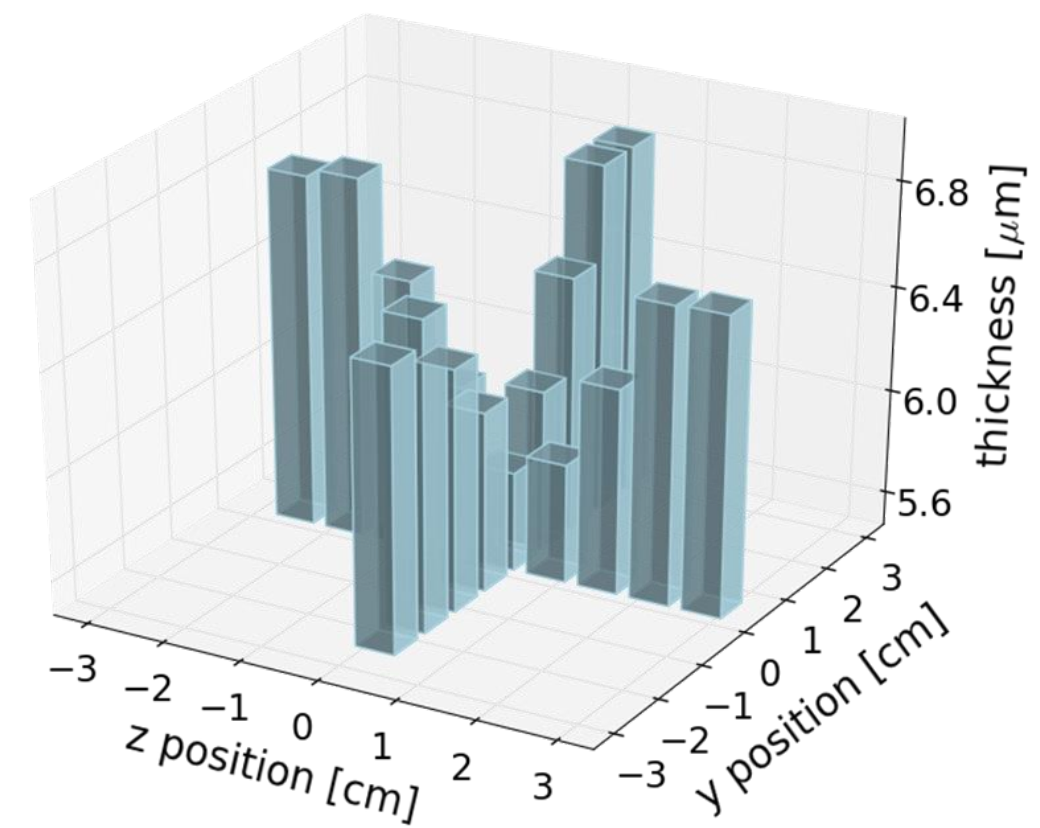} 
\caption{The thickness measured at different positions of the LN wafer varied from roughly $4L_c$, at the center, to about $5L_c$ at the edge.}
\label{fig:thickness}
\end{figure}


\begin{figure}[htb]
\includegraphics[width=\columnwidth,trim={0 9cm 1cm 9cm},clip]{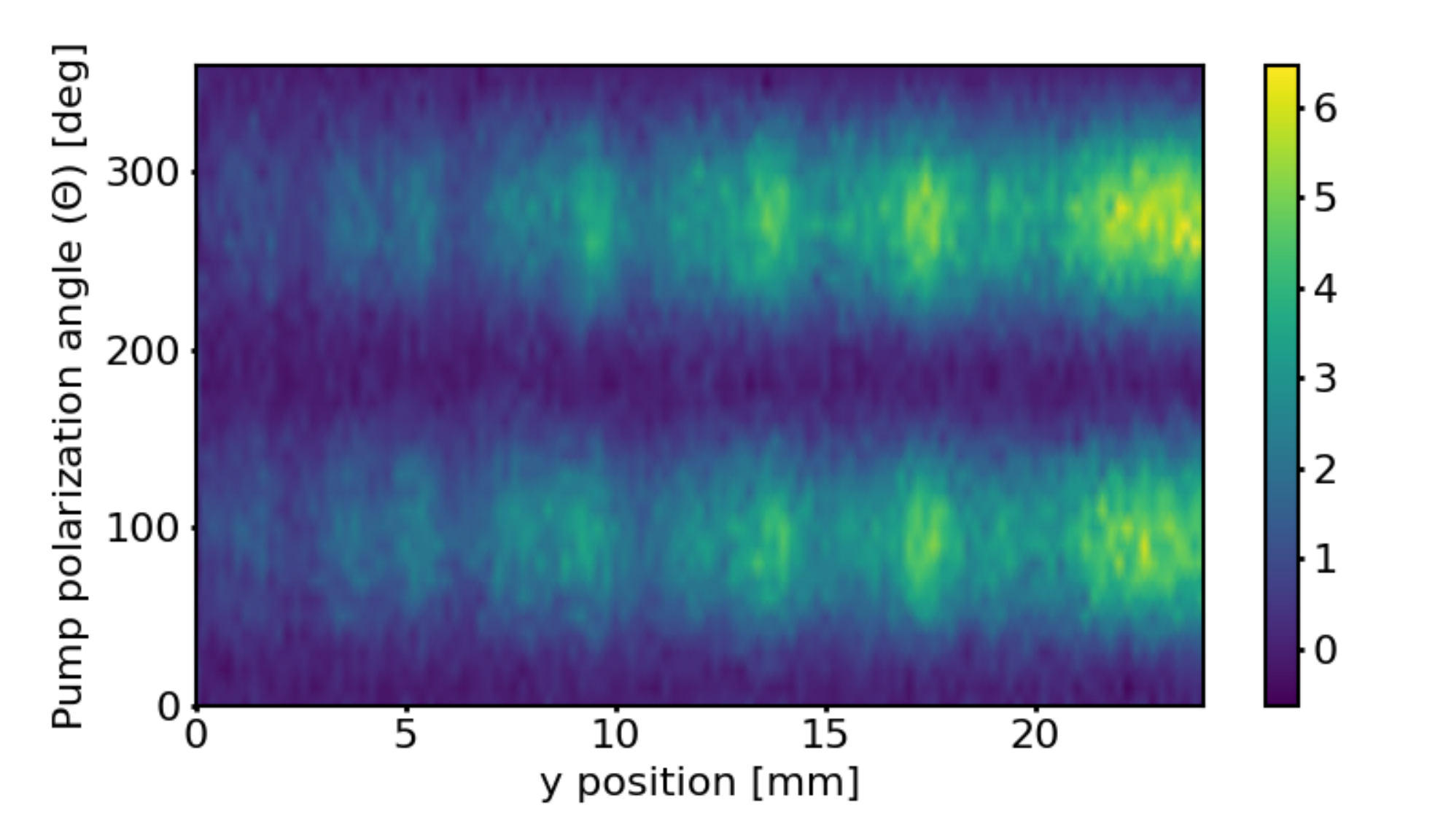} 
\caption{Coincidence rate for z-polarized emission as a function of the y position and the pump polarization. At position '0' the crystal thickness corresponds to about 4$L_c$ and therefore no SPDC is expected. On the contrary, at the edge of the crystal its thickness is about 5$L_c$ and therefore we expect the maximum emission as confirmed by the increased coincidence rate. Coincidences and therefore SPDC are present only when the pump is polarized along the z axis (which corresponds to 90 and 270 degrees when the polarisation angle is measured from the y axis).}
\label{fig:2DCoincMap}
\end{figure}

\begin{figure}[htb]
\includegraphics[width=\columnwidth]{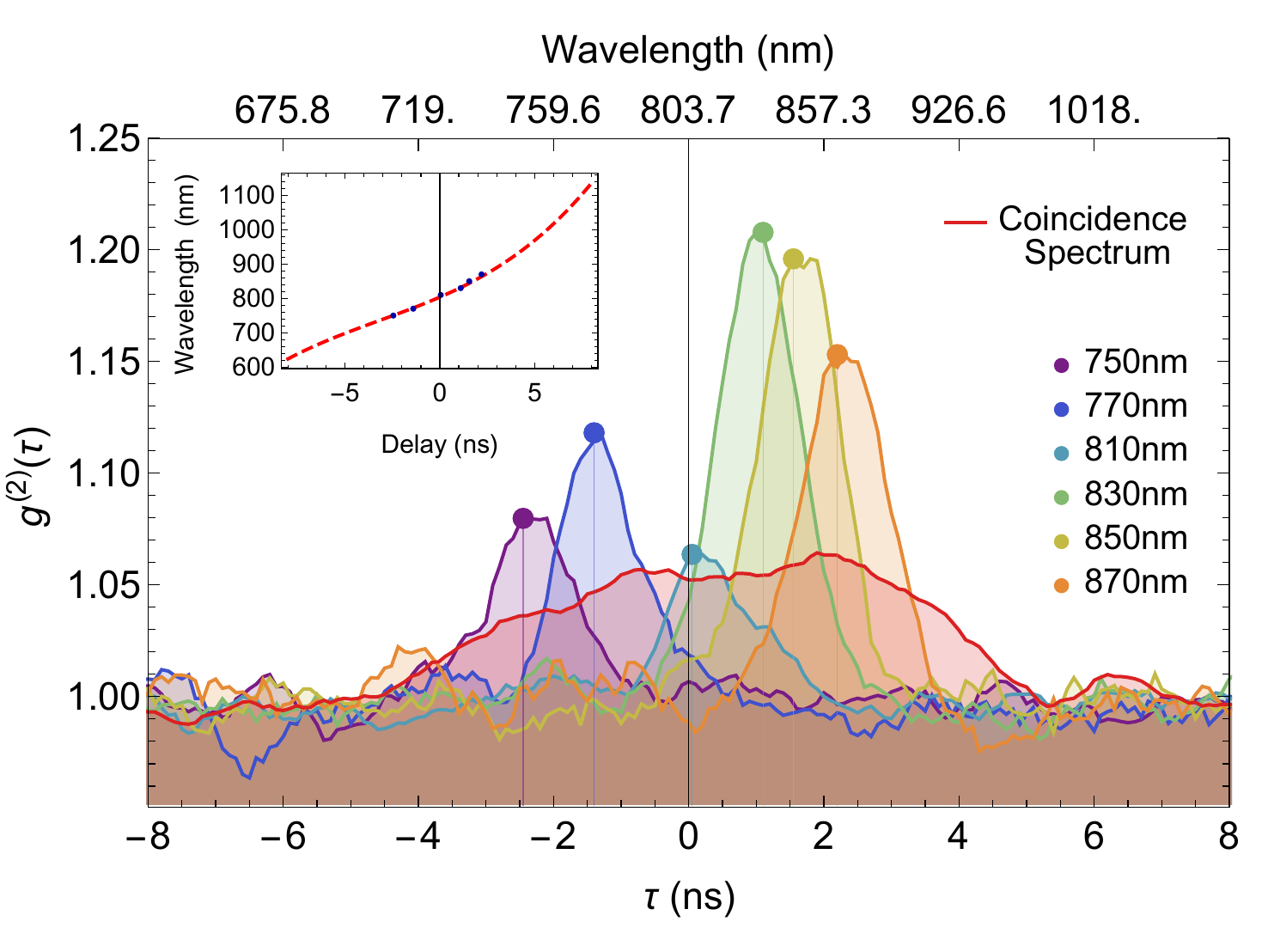} 
\caption{The normalised correlation function for different time delays for the SPS measurement. Each narrow peak corresponds to a calibration wavelength and the total coincidence spectrum, broadened due to the propagation through the fiber, is shown in red. The inset shows the calibration curve to map time differences to wavelength differences. The experimentally found blue points are fitted with a cubic polynomial in red.}
\label{fig:calibration}
\end{figure}

\begin{figure}[htb]
\includegraphics[width=\columnwidth]{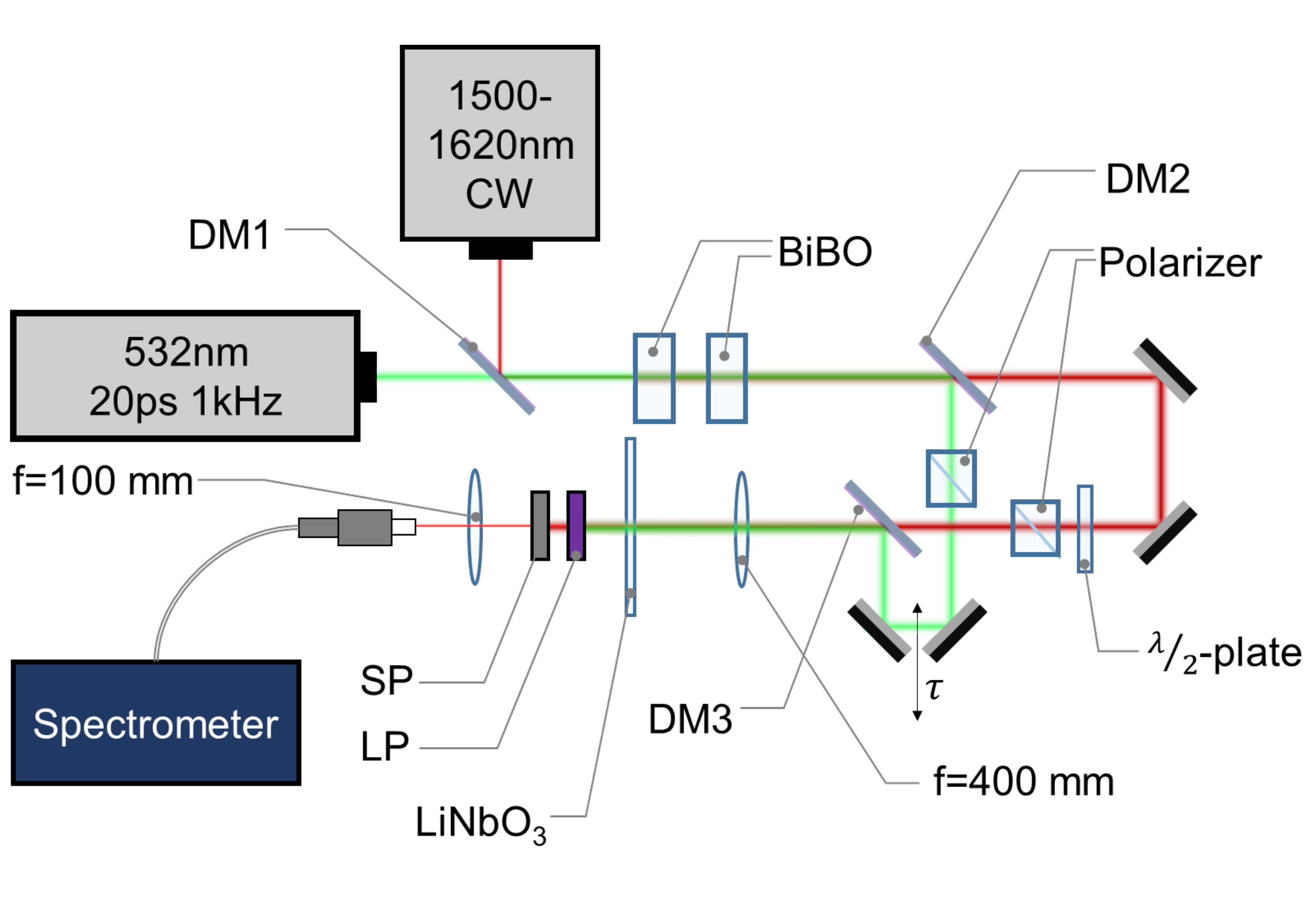} 
\caption{A schematic of the setup used to reconstruct the JSI using SET. DM1,2 and 3 denote dichroic mirrors that spilt 532nm and IR wavelengths. SP and LP denote short and long pass filters to remove the pump and seed wavelengths. $\tau$ is the delay line that compensates for the temporal walk-off between the pump and seed beams.}
\label{fig:setupset}
\end{figure}

\end{document}